\documentclass[prd,twocolumn,nofootinbib]{revtex4}

\usepackage{graphicx}
\usepackage{dcolumn}
\usepackage{amsmath}
\usepackage{epsfig}
\usepackage{amsfonts}
\usepackage{amssymb}
\usepackage{revsymb}

\newcommand{\uvec}[1]{\widehat{#1}}

\makeatother

\begin{document}

\title{\bf Relativistic effects of our galaxy's motion on circles--in--the--sky }

\author{M.O. Calv\~ao}
\email{orca@if.ufrj.br}
\affiliation{Universidade Federal do Rio de Janeiro\\
Instituto de F\'{\i}sica, C.P. 68528 \\
21945-972 Rio de Janeiro -- RJ, Brazil}

\author{G.I. Gomero}
\email{german@ift.unesp.br}
\affiliation{Instituto de F\'{\i}sica Te\'orica \\
Universidade Estadual Paulista \\
Rua Pamplona 145 \\
01405-900 S\~ao Paulo -- SP, Brazil \\}

\author{B. Mota}
\email{brunom@cbpf.br}

\author{M.J. Rebou\c{c}as}
\email{reboucas@cbpf.br}
\affiliation{Centro Brasileiro de Pesquisas F\'{\i}sicas\\
Departamento de Relatividade e Part\'{\i}culas \\
Rua Dr.\ Xavier Sigaud 150 \\
22290-180 Rio de Janeiro -- RJ, Brazil}
\date{\today}

\begin{abstract}
For an observer in the Hubble flow (comoving frame) the last
scattering surface (LSS) is well approximated by a two-sphere. If
a nontrivial topology of space is detectable, then this sphere
intersects some of its topological images, giving rise to
circles-in-the-sky, i.e., pairs of matching circles of equal
radii, centered at different points on the LSS sphere, with the
same pattern of temperature variations. Motivated by the fact that
our entire galaxy is not exactly in the Hubble flow, we study the
geometric effects of our galaxy's peculiar motion on the
circles-in-the-sky. We show that the shape of these
circles-in-the-sky remains circular, as detected by a local
observer with arbitrary peculiar velocity. Explicit expressions
for the radius and center position of such an observed
circle-in-the-sky, as well as for the angular displacement of
points on the circle, are derived. In general, a circle is
detected as a circle of different radius, displaced relative to
its original position, and centered at a point which does not
correspond to its detected center in the comoving frame. Further,
there is an angular displacement of points on the circles. These
effects all arise from aberration of cosmic microwave
background radiation, exhausting the purely geometric
effects due to the peculiar motion of our galaxy, and are
independent of both the large scale curvature of space and the
expansion of the universe, since aberration is a purely local
phenomenon. For a Lorentz-boosted observer with the speed of our
entire galaxy, the maximum (detectable) changes in the angular
radius of a circle, its maximum center displacement, as well as
the maximum angular distortion are shown all to be of order
$\beta=(v/c)$ radians. In particular, two back-to-back matching
circles in a finite universe will have an upper bound of $2|\beta|
$ in the variation of either their radii, the angular position of
their centers, or  the angular distribution of points.
\end{abstract}


\maketitle

\section{Introduction}
\label{sec:intro}
Whether the universe is spatially finite and what its size and shape
may be are among the fundamental open problems that high precision
modern cosmology seeks to resolve. These questions of topological
nature have become particularly topical, given the wealth of
increasingly accurate cosmological observations, especially
the recent results from the Wilkinson Microwave Anisotropy Probe
(WMAP) experiment~\cite{WMAP}, which have heightened the interest
in the possibility of a finite universe. Indeed, reported non-Gaussianity
in cosmic microwave background (CMB) maps~\cite{CMB+NonGauss}, the
small power of large-angle fluctuations~\cite{WMAP-Spergel-et-al},
and some features in the power spectrum~\cite{CMB+NonGauss,WMAP-Spergel-et-al}
are large-scale anomalies which have been suggested as potential indication
of a finite universe~\cite{Poincare} (for reviews see~\cite{Revs}).

Given the current high quality and resolution of such maps, the most
promising search for cosmic topology is based on pattern repetitions
of these CMB anisotropies on the last scattering surface (LSS). If a
nontrivial topology of space is detectable%
\footnote{The extent to which a nontrivial topology may or may not
be detected has been discussed in references~\cite{TopDetec}.},
then the last scattering sphere intersects some of its
topological images giving rise to the so-called circles-in-the-sky.
Thus, the CMB temperature anisotropy maps will have matched
circles: pairs of equal radii circles (centered at different
points on the LSS sphere) that have the same
pattern of temperature variations~\cite{CSS1998}.
These matching circles will exist in CMB anisotropy maps of universes
with any detectable nontrivial topology, regardless of its geometry.
Therefore, pairs of matched circles may be hidden in CMB maps if
the universe is finite, and to observationally probe nontrivial
topology on the largest scale available, one needs a statistical
approach to scan all-sky CMB maps in order to draw the correlated
circles out of them. The circles-in-the-sky method, devised by
N. Cornish, D. Spergel and G. Starkman~\cite{CSS1998} to search for
a possible nontrivial topology of the universe, looks for such
matching circles through a correlation statistic for sign
detection (a function whose peaks indicate matched circles).

As originally conceived, the circles--in--the--sky method did not
take into account the role of our galaxy's peculiar motion. In a
recent paper~\cite{Levin2004}, however, this point has been
considered, and it has been argued, in a simplified context (flat
spacetime), that, for any observer moving with respect to the CMB,
two effects will take place, namely the circles will be deformed
into ovals, and these ovals will be displaced with regard to the
corresponding circles in the comoving frame. These effects were
estimated to be, respectively, of order $\beta^2$ and $|\beta|$.

We show here that, regardless of any background curvature or
expansion, the shape of these circles--in--the--sky, as locally
detected by an observer in motion relative to the comoving one,
\emph{remains} circular. We derive explicit expressions for the
radius and center position of such an observed circle-in-the-sky,
as well as a formula for the angular displacement of points on the
circle. In general, a circle is detected as a circle of different
radius, displaced relative to its original position, and centered
at a point which does not correspond to its detected center in the
comoving frame. Further, there is an angular displacement of
points on the circles. These effects arise all from aberration of
CMBR, and exhaust the purely geometric effects due to our galaxy's
peculiar motion on circles-in-the-sky. We also estimate the
maximum values of these effects considering the peculiar motion of
our galaxy. In particular, for two back-to-back matching circles
in a finite universe the upper bounds in the variation of either
their radii, the angular position of their center or in the
angular distribution of points are all of order $2|\beta|\simeq
2.46\times 10^{-3}$ radians. Although these effects are still
below WMAP's angular resolution, we show that they are relevant
for future CMB missions like the Planck satellite.

\section{Main Results}
\label{sec:MainRes}

We begin by recalling an old beautiful result by R. Penrose%
~\cite{Penrose1959}, according to which, if an observer $O$ detects,
in  his sky-sphere, the shape of an object as circular, then for any
other observer $O'$ locally coinciding with and in motion relative
to $O$, the detected shape will remain circular. The gist of
the reasoning is based on the relativistic aberration and the
stereographic projection properties, as described in more detail by W. Rindler~\cite{Rindler}: the outline of a spherical object, projected
onto the local celestial sphere (``sky'') of a given observer
$O$ is certainly circular and, by means of a suitable stereographic
projection, is mapped onto a circle in the projection plane (``screen'').
Now, the aberration formula just transforms this screen for $O$ onto a
corresponding screen for $O'$ which is merely globally expanded
(or contracted) by a factor $[(1-\beta)/(1+\beta)]^{1/2}$; by the
corresponding inverse stereographic projection, we thus arrive at the
above mentioned Penrose's result. As a consequence, circles-in-the-sky as
detected by the observer $O$ will also be detected as circles-in-the-sky
by the observer $O^\prime$.
The circles will, in general, differ both in position and in size as
we shall discuss in detail in what follows.

Let $P^\mu = (h\nu/c) (1,-\uvec{n})$ be the $4$-momentum of an
incoming photon in a direction $\uvec{n}$ and with frequency $\nu$
as detected by an observer $O$ in the Hubble flow (comoving
observer). For another observer $O^\prime$ who moves with velocity
$\vec{\beta}$ relative to this comoving observer, and whose
spatial position coincides with that of $O$ at the time they
measure the CMB, the photon will be detected as incoming in a
different direction $\uvec{n}'$, and with a distinct frequency
$\nu'$. Clearly, the $4$-momentum of this photon for the observer
$O'$ is given by the Lorentz transformation
\begin{equation}
\label{NewP}
P^{\prime\mu} = \Lambda^\mu_{\ \,\sigma} P^\sigma \; ,
\end{equation}
where
\begin{equation}
\label{Lorentz}
\Lambda^\mu_{\ \,\sigma} = \left( \begin{array}{cc}
                     \gamma &  - \gamma \vec{\beta}^T \\
                   - \gamma \vec{\beta} &  \;\;\,I + (\gamma - 1) \,
                                      \uvec{\beta} \, \uvec{\beta}^T
                         \end{array} \right) \;.
\end{equation}
Here $I$ stands for the $3$-dimensional identity matrix,
$\gamma = (1-\beta^2)^{-1/2}$, $\vec{\beta}^T$ it the
transpose of $\vec{\beta}$, and $\uvec{\beta}=\vec{\beta}/\beta$.
Thus we readily obtain
\begin{eqnarray}
\label{Doppler}
\nu' & = & \gamma \left( 1 + \vec{\beta} \cdot \uvec{n} \right) \nu
\; , \\
\label{Aber}
\uvec{n}' & = & \frac{\uvec{n} + \left[ (\gamma - 1) \uvec{\beta} \cdot
\uvec{n} + \gamma \beta \right] \uvec{\beta}}{\gamma \left( 1 + \vec{\beta}
\cdot \uvec{n} \right)} \; .
\end{eqnarray}

Equation~(\ref{Doppler}) gives the Doppler effect, i.e. a shift in
frequency due to our galaxy's peculiar motion, responsible for the
large dipole moment in the CMB temperature anisotropies and for
tiny spectral distortions in the corresponding maps.
Equation~(\ref{Aber}), on the other hand, expresses the relativistic
aberration of light, which gives rise to the position displacement
and re-scaling  as well as the angular displacement of points
of the circles-in-the-sky due to the motion of $O'$ relative to $O$.

We first use~(\ref{Aber}) to determine the center and radius of a
Lorentz-boosted circle-in-the-sky. Without loss of generality, let
us choose the common axes $z$ and $z'$ such that their positive
direction coincides with the direction of the velocity $\vec{\beta}$.
Thus, from~(\ref{Aber}), the transformation of the direction of
the incoming photon, $(\theta,\varphi) \mapsto (\theta',\varphi')$,
reduces to
\begin{equation}
\label{AberEsf}
\cos \theta' =\frac{\beta + \cos \theta}{1 + \beta \cos \theta}
\qquad \mbox{and} \qquad \varphi' = \varphi \;,
\end{equation}
where $\theta$ and $\varphi$ are the usual spherical coordinates.

Let $\uvec{q}_c = (\theta_c, \varphi_c)$ and $\rho$ be, respectively,
the center and radius of a circle $C(\uvec{q}_c,\rho)$ in the
comoving frame. We note that the direct use of~(\ref{AberEsf}) to
the center $\uvec{q}_c$ of $C$ does not furnish, in general, the
center $\uvec{q}^\prime_c$ of the Lorentz-boosted circle
$C'(\uvec{q}^\prime_c,\rho')$,  because stereographic projection
of a circle $C$ is a circle whose center is not, in general, the
stereographic projection of the center of $C$.
Therefore, in order to calculate the expressions for the center
$\uvec{q}^\prime_c = (\theta^\prime_c, \varphi^\prime_c)$ and the
radius $\rho'$ of the Lorentz-boosted circle $C'$ we use~(\ref{AberEsf})
to transform the top and bottom points of the circle $C$, given
by $(\theta_c - \rho, \varphi_c)$ and $(\theta_c + \rho, \varphi_c)$,
respectively. As  $\varphi' = \varphi$, obviously one has
$\varphi_c' = \varphi_c$, and the transformed points remain,
respectively, the top and bottom points of the Lorentz-boosted
circle $C'$.
{}From~(\ref{AberEsf}) we have
\begin{equation}
\label{TopBPoint}
\cos(\theta^\prime_c - \rho') = \frac{\beta + \cos(\theta_c - \rho)}{1 +
\beta \cos(\theta_c - \rho)}
\end{equation}
for the top point, and a similar expression for $(\theta_c' + \rho')$.
Now using some elementary trigonometric relations, after some calculation
we obtain
\begin{equation}
\label{BoostedCircle}
\sin \theta^\prime_c = \frac{1}{\gamma \, M} \, \sin \theta_c \;,
\quad \mbox{and} \quad
\sin \rho^\prime = \frac{1}{\gamma \,M} \, \sin \rho \; ,
\end{equation}
where
\begin{equation}
\label{M}
M = \sqrt{ [1+\beta \cos(\theta_c + \rho)]
                       [1+\beta \cos(\theta_c-\rho)] } \; .
\end{equation}

We emphasize that, by using ~(\ref{AberEsf}) and~(\ref{BoostedCircle}),
it can readily be verified that the circle-in-the-sky $C(\uvec{q}_c,\rho)$
transforms into the circle-in-the-sky $C'(\uvec{q}'_c,\rho')$,
i.e., the circle equation
\begin{equation} \label{CircleEq}
       \uvec{q}\cdot\uvec{q}_c = \cos\rho
\end{equation}
\noindent is invariant under the transformation given by~(\ref{AberEsf})
and~(\ref{BoostedCircle}). In (\ref{CircleEq})
$\uvec{q}$ stands for an arbitrary point in the circle (cf.
eq.(\ref{Decomp}) below).

Now, for non-relativistic velocities ($|\beta| \ll 1$), at first order
in $\beta$ we have
\begin{eqnarray}
\label{NewPos}
\theta^\prime_c & = & \theta_c - \beta \sin \theta_c \cos \rho \;,
\\
\label{NewRadius}
    \rho' & = & \rho - \beta \cos \theta_c \sin \rho \; .
\end{eqnarray}

\begin{figure}[!tbh]
\centerline{\def\epsfsize#1#2{0.5#1}\epsffile{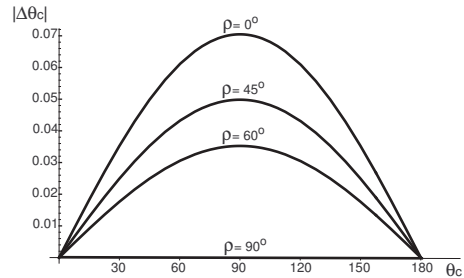}}
\caption{\label{fig:DelThe} The behavior of displacement $|\Delta \theta_c|$
of the position of the center for different radii of the circles. In this
figure $\beta = 1.23 \times 10^{-3}$. }
\end{figure}

It follows from~(\ref{NewPos}) that for a given circle of radius
$\rho$ the displacement $|\Delta \theta_c|= |\theta_c'-\theta_c|$
of its center is maximum for $\theta_c=\pi/2$, i.e. when the vector position
of its center in the comoving frame is orthogonal to the boost-velocity
$\vec{\beta}$. In this case the angular radius $\rho$ remains unchanged
as detected by $O'$. The upper bound of  $|\Delta \theta_c|$ clearly
is $|\beta|$. Thus, two back-to-back circles-in-the-sky in a universe
with a nontrivial topology will have a maximum displacement of
$2|\beta|$ in the position of their centers.
Figure~\ref{fig:DelThe} shows, for a fixed $\beta$, the behavior of
$|\Delta \theta_c|$ as a function of $\theta_c$ for circles of
different radii. It is apparent from this figure, that the maximum
absolute value of the angular displacement of the center of the
circles depends on their radius but occurs for $\theta_c=\pi/2$,
while for a circle of angular radius $\rho=\pi/2$ there is no change
in the position of the center no matter what the center position is.

\begin{figure}[!thb]  
\centerline{\def\epsfsize#1#2{0.5#1}\epsffile{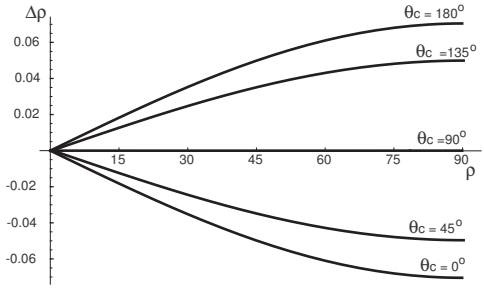}}
\caption{\label{fig:DelRho1} The behavior of displacement $\Delta
\rho$ of the radius of the circles as a function of $\rho$ for
different locations of the center of the circles. In this
figure $\beta = 1.23 \times 10^{-3}$.}
\end{figure}

\begin{figure}[!tbh]
\centerline{\def\epsfsize#1#2{0.4#1}\epsffile{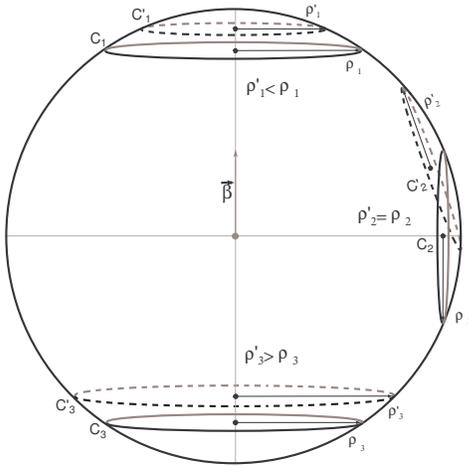}}
\caption{\label{fig:CinSk2.eps} A schematic illustration of the
changes in the radii and in the positions of the center of
the circles. The continuous and dashed circles indicate
circles-in-the-sky as detected, respectively, by the observers $O$
and $O'$. The circles $C_2$ and $C_2'$ illustrate the maximum
change in the position of the center. In this case the radius
remains the same. The circles $C_1$ and $C_1'$
represent the instance in which the radius of the circle has
a maximum decrease, while $C_3$ and $C_3'$ illustrate the case
when the radius has a maximum increase. The set of circles
$\{(C_1,C_3), (C_1',C_3')\}$ stands for the situation of
maximum change in back-to-back matching circles-in-the-sky.}
\end{figure}

On the other hand, for a given circle $C$ of angular radius $\rho$ the
change $\Delta \rho = \rho'- \rho$ is maximum when the direction of
vector position of the center of $C$ in the comoving frame is the
same direction of $\vec{\beta}$, i.e. $\theta_c=0$ or $\theta_c = \pi$.
In each of these cases the position of the center of the circle as
detected by $O'$ remains fixed.
We note that when $\theta_c=0$ the radius of the circle decreases as
detected by the observer $O'$, while for $\theta_c = \pi$ it increases
(see also figure~\ref{fig:CinSk2.eps}).
The upper bound for $|\Delta \rho|$ clearly is again $|\beta|$.
Figure~\ref{fig:DelRho1} shows the behavior of $\Delta \rho$ as
a function of $\rho$ for distinct positions of the centers
$\theta_c$. It is clear from this figure that circles whose vector
position of the center is orthogonal to the velocity $\vec{\beta}$
do not change the radii, while for circles whose vector position
is parallel to (antiparallel) $\beta$ we have a maximum absolute
variation of $\rho$ for any given fixed circle. Clearly for
$0^\circ \leq \theta_c < 90^\circ$,  $\Delta \rho$ is
a decreasing function of $\rho$, while for $90^\circ < \theta_c
\leq 180^\circ$ it is an increasing function of $\rho$.

{}From ~(\ref{NewRadius}) we have that for $\theta_c= 90^\circ$
the radius remains unchanged.
The upper bound for $|\Delta \rho|$ obviously is $|\beta|$.
Thus, two back-to-back correlated circles in a finite universe with
a detectable topology will have a upper bound of $2|\beta|$ in the
variation of the radii as detected by the moving observer $O'$.

Figure~\ref{fig:CinSk2.eps} shows  the changes in the radii
and in the positions of the center of the circles in three
instances. The continuous and dashed circles indicate
circles-in-the-sky as detected, respectively, by the observers
$O$ and $O'$. The circles $C_2$ and $C_2'$ correspond to a
case of maximum change in the position of the center. Clearly
in this case the radius $\rho$ remains unchanged at first order
in $\beta$.
The circles $C_1$ and $C_1'$ correspond to a case in which
the radius of the circle decreases, while $C_3$ and $C_3'$
shows a case when the radius increases. In these two last cases
the positions of the center of the circles remain unchanged.
The pairs of matching circles $(C_1,C_3)$ and $(C_1',C_3')$
illustrate the instance of maximum change in back-to-back
correlated circles-in-the-sky.

To fully characterize all geometric effects which arise from
aberration of CMB on circles-in-the-sky we shall consider now the
angular displacement of points on a circle as detected by the
observers $O$ and $O'$. To this end, we decompose a point
$\uvec{q} = (\theta, \varphi)$ on a circle $C(\uvec{q}_c,\rho)$ as
\begin{equation}
\label{Decomp} \uvec{q} = \vec{p} + \uvec{q}_c \cos \rho \; ,
\end{equation}
where $\vec{p}$ is on the plane of the circle, which is orthogonal
to $\uvec{q}_c = (\theta_c, \varphi_c)$. The vector $\vec{p}$ is
the orthogonal complement of the projection of $\uvec{q}$ onto
$\uvec{q}_c$. We take $\vec{p}_0$, the orthogonal complement of
the top point $\uvec{q}_0 = (\theta_c - \rho, \varphi_c)$ of $C$,
as a reference from which we measure the angle $\alpha$ defined by
\begin{equation}
\label{alphaDef}
\cos \alpha = \frac{\vec{p} \cdot \vec{p}_0}{\sin^2 \rho} \;,
\end{equation}
since clearly $|\vec{p}|= |\vec{p}_0|= \sin \rho$.

{}From~(\ref{Decomp}) one easily has
\begin{equation}
\label{alpha1}
\cos \alpha = \frac{\cos\theta \sin\theta_c - \sin\theta \cos\theta_c
\cos (\varphi - \varphi_c)}{\sin \rho} \; ,
\end{equation}
or using the circle equation~(\ref{CircleEq}) to eliminate the dependence
on $\varphi - \varphi_c$,
\begin{equation}
\label{alpha2}
\cos \alpha = \frac{\cos \theta - \cos \theta_c \cos \rho}{\sin \theta_c
\sin \rho} \; .
\end{equation}
To calculate $\cos \alpha^\prime$ it is more convenient to substitute
(\ref{AberEsf}) and (\ref{BoostedCircle}) into the analog of~(\ref{alpha1}),
for $\cos \alpha^\prime$ (which clearly must hold) and then
use~(\ref{CircleEq}) and~(\ref{alpha2}) to simplify the expression to
\begin{equation}
\label{alphaprime}
\cos \alpha^\prime =\cos\alpha + \left( \frac{\beta}{1 + \beta \cos\theta}
\right) \, \sin \rho \sin \theta_c \sin^2 \alpha \; ,
\end{equation}
where
$\cos \theta = \cos\rho\ \cos\theta_c + \sin\rho\ \sin\theta_c \cos \alpha$.
At first order in $\beta$ equation~(\ref{alphaprime}) becomes
\begin{equation}
\label{alphaprime2}
\alpha^\prime = \alpha - \beta \sin \rho \sin \theta_c \sin \alpha \; .
\end{equation}

\begin{figure}[!tbh]
\centerline{\def\epsfsize#1#2{0.4#1}\epsffile{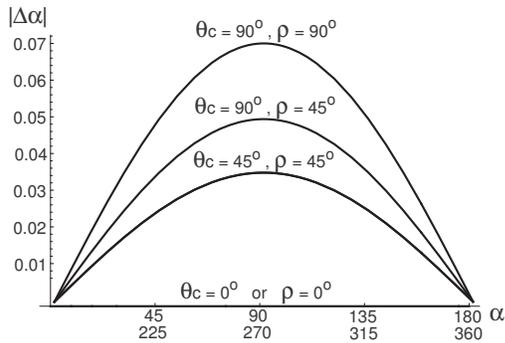}}
\caption{\label{fig:DelAlp} The behavior of angular displacement
$|\Delta \alpha|$ for circles of different radii and centers. In
this figure $\beta = 1.23 \times 10^{-3}$. }
\end{figure}

It follows from~(\ref{alphaprime2}) that for a given circle of radius
$\rho$ the angular displacement $|\Delta \alpha|$ is maximum
for $\theta_c = \pi/2$ and $\alpha = \pm \pi/2$.
The upper bound for $|\Delta \alpha|$ clearly is $|\beta|$ and
occurs when simultaneously $\theta_c = \pi/2$, $\rho = \pi/2$ and
$\alpha=\pi/2$.
Here again, two back-to-back matching circles in a multiply connected
universe will have a upper bound of $2|\beta|$ in the variation of the
angular displacement.
Figure~\ref{fig:DelAlp} shows the behavior of $|\Delta \alpha|$ as
a function of $\alpha$ for different values of the radii and location
of the center of circles. It is clear from this figure that for any
given circles  center at 
$\theta_c \neq 0^\circ$ the value of maximum angular distortion
depend on the radii and centers of the circles but takes place
for $\alpha = \pm \pi/2$, while for circles with $\rho$
whose vector position of the centers are parallel (or antiparallel) to
$\vec{\beta}$ ($\theta_c=0^\circ$ or $\theta_c = \pi $) there is no
angular distortion.

\begin{figure}[!tbh]
\centerline{\def\epsfsize#1#2{0.41#1}\epsffile{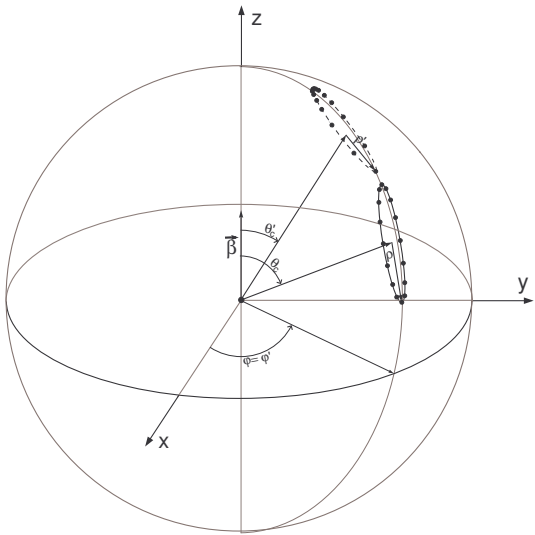}}
\caption{\label{fig:CinSkPhase} A schematic illustration of the
angular displacement effect. It shows how a set of points
homogeneously distributed on the solid line circle
$C(\uvec{q}_c,\rho)$ as detected by an observer $O$, is detected
by the observer $O'$ in the new dashed circle
$C'(\uvec{q}_c',\rho')$.}
\end{figure}

\begin{figure}[!tbh]
\centerline{\def\epsfsize#1#2{0.45#1}\epsffile{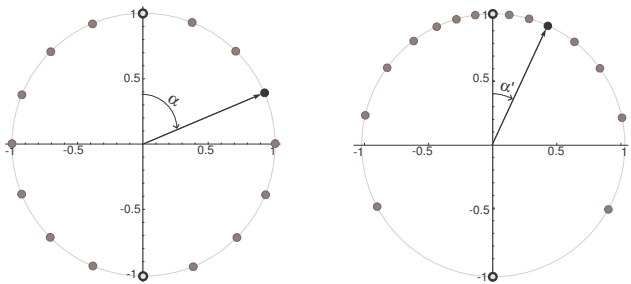}}
\caption{\label{fig:AngPhase} Points regularly distributed along a
circle (left), as detected by observer $O$, are distorted by
aberration when detected by observer $O'$ (right). The
points lying on the meridian ($\alpha$=0 and $\alpha=180^{\circ}$)
are unaffected by angular distortion. In this figure the exact
expression (\ref{alphaprime}) was used with $\beta=0.8$.}
\end{figure}

Figures~\ref{fig:CinSkPhase} and \ref{fig:AngPhase} illustrate the
angular distortion effect. In these figures it is shown how a set of
points uniformly distributed along the circle $C(\uvec{q}_c,\rho)$
as detected by the comoving observer $O$ will be detected by a
locally coinciding observer $O'$ in the transformed circle
$C'(\uvec{q}_c',\rho')$.

\section{Concluding Remarks}
\label{ConRem}

Motivated by the fact that our entire galaxy is not in the Hubble
flow, we have studied all geometric effects due to our galaxy's
peculiar motion on the circles-in-the-sky. As a consequence of
Peronse's result~\cite{Penrose1959} we have shown that circles-in-the-sky
as detected by a comoving observer $O$ will also be detected as
circles-in-the sky by any observer $O'$ locally coinciding
with $O$ at the time they measure the CMBR.
In general, a circle for $O$ will be detected by $O'$ as a circle of
different radius, displaced relative to its original position, and
centered at a point which does not correspond to its detected center
in the comoving frame.

We have derived closed explicit expressions for the radius, $\rho'$,
the center position, $(\theta_c', \phi_c')$, and the angular displacement
$\alpha'$ of points on the circles as detected by the Lorentz-boosted
observer $O'$. We have also plotted figures to illustrate several
instances of these effects.

The maximum displacement in either radius or center or
angular position of points on a circle is $\pm \beta$. From the
value $\beta=1.23~\times~10^{-3}$  obtained from the dipole
amplitude in the CMB spectrum, we have that the maximum
displacement for each of these effects individually is $\simeq \pm
0.07^{\circ}=4.2^{\prime}$. Thus, two antipodal circles whose
center vector positions  are parallel (antiparallel) to
$\vec{\beta}$ will have a difference in radii of $0.14^{\circ}$,
as detected by $O$ and $O'$. This is below the current WMAP
resolution (at best $0.25^{\circ}= 15'$)~\cite{WMAP}, but close to
Planck's (at least $0.16^{\circ}=10^{\prime}$)~\cite{Planck} and
other more accurate forthcoming missions' resolution.

In Ref.~\cite{Levin2004}, in a simplified flat spacetime, the
spatial positions, relative to a ``moving'' frame, of the events
defined by the intersection of the copies of the LSS spheres were
determined; of course, these events are simultaneous in the CMB
frame and, \emph{ipso facto}, are non-simultaneous in the
``moving'' one. The specific procedure of collecting the spatial
positions of these \emph{non-simultaneous} events defines, in the
``moving'' frame, an oval figure, whose diameter along the boost
direction is \emph{Lorentz-dilated}, whereas its perpendicular
diameter remains unchanged. Briefly, it was shown that a (spatial)
circle in the CMB frame, as a geometric figure, is formally
transformed into a (spatial) oval in the ''moving'' frame of the
underlying flat spacetime. The approach we adopted to the effects
of our galaxy's peculiar motion on the circles-in-the-sky is
realistic and complete, since it does not rely on such a flat
spacetime, and takes into account that a typical observation of
the CMBR records simultaneously incoming light rays in an
essentially infinitesimal detector used by a local observer, thus
implying their projection onto the sky-sphere.

We note that the aberration of CMBR might be of a more general
interest in the analysis of maps of CMB temperature anisotropies.
Indeed, consider the celestial sphere pixelized by a comoving
observer $O$ and consider another observer $O'$ coinciding with
$O$ and with relative  velocity $\vec{\beta}$ along the
positive $z$ direction. The coordinates $(\theta_i, \varphi_i)$ of
the $i$--th pixel in the comoving frame are transformed to the
moving frame according to~(\ref{AberEsf}), which, at first order
in $\beta$, simply reads
\begin{equation}
\label{AberEsfNR}
\theta^\prime_i = \theta_i - \beta \sin \theta_i \qquad ; \qquad
\varphi^\prime_i = \varphi_i \; .
\end{equation}
This displacement of the centers of the pixels gives rise to a
distortion on the temperature variation pattern in CMB maps,
which might be relevant for future missions.

Finally, we emphasize that all these effects arise from aberration
of the CMBR, and exhaust the purely geometric effects due to
the motion of our galaxy (or CMBR detectors). It is again
worth noting that, since aberration of light is a purely
local phenomenon, the effects considered in this work depend
neither on the background curvature of space nor on the expansion
of the universe.

\bigskip
\noindent{\bf Acknowledgments}
\medskip

We thank CNPq and FAPESP (contract 02/12328-6) for the grants under
which this work was carried out.

\end{document}